# Spatiotemporal Traffic Prediction in Distributed Backend Systems via Graph Neural Networks

Zhimin Qiu
University of Southern California
Los Angeles, USA

Feng Liu
Stevens Institute of Technology
Hoboken, USA

Yuxiao Wang
University of Pennsylvania
Philadelphia, USA

Chenrui Hu
University of Pennsylvania
Pennsylvania, USA

Ziyu Cheng
University of Southern California
Los Angeles, USA

Di Wu*
Washington University in St. Louis
St. Louis, USA

*Abstract-This paper addresses the problem of traffic prediction in distributed backend systems and proposes a graph neural network – based modeling approach to overcome the limitations of traditional models in capturing complex dependencies and dynamic features. The system is abstracted as a graph with nodes and edges, where node features represent traffic and resource states, and adjacency relations describe service interactions. A graph convolution mechanism enables multi-order propagation and aggregation of node features, while a gated recurrent structure models historical sequences dynamically, thus integrating spatial structures with temporal evolution. A spatiotemporal joint modeling module further fuses graph representation with temporal dependency, and a decoder generates future traffic predictions. The model is trained with mean squared error to minimize deviations from actual values. Experiments based on public distributed system logs construct combined inputs of node features, topology, and sequences, and compare the proposed method with mainstream baselines using MSE, RMSE, MAE, and MAPE. Results show that the proposed method achieves stable performance and low error across different prediction horizons and model depths, significantly improving the accuracy and robustness of traffic forecasting in distributed backend systems and verifying the potential of graph neural networks in complex system modeling.*

## I. INTRODUCTION

In the context of rapid informationization and digitalization, backend systems play a crucial role in handling massive requests, allocating computing resources, and ensuring service continuity[1]. With the widespread adoption of cloud computing, the Internet of Things [2], and large language models platforms [3-5], backend traffic demonstrates high dynamics and complexity. It contains both periodic and bursty variations and is further influenced by user behavior patterns, application load characteristics, and network fluctuations. Effective prediction of backend system traffic has become a fundamental requirement for ensuring service stability, improving scalability, and optimizing resource allocation. Accurate traffic forecasting enables early identification of potential congestion risks and supports resource scheduling, cache allocation, and load balancing.

Traditional methods often rely on statistical modeling or single time series analysis[6]. These approaches face clear limitations when dealing with complex nonlinear dependencies and multi-dimensional interactions. The operational data of distributed backend systemsss is not isolated. They are interconnected through network topology, node dependencies, and service invocation relationships, forming a complex structure. Relying solely on historical traffic curves is insufficient to capture such structured information or to describe the interdependence and propagation effects among nodes. This highlights the need for models that can process graph-structured data, thereby better capturing node – edge interactions and enhancing the understanding and prediction of global system traffic evolution[7].

The emergence of graph neural networks provides a new opportunity to address this challenge. Graph neural networks propagate and aggregate information based on graph topology and node features, allowing the extraction of spatiotemporal dependencies at multiple levels and dimensions[8]. Compared with traditional methods, they can preserve local features while integrating global associations. This offers both higher accuracy and better interpretability in modeling distributed systems. Such an approach not only improves prediction at individual nodes but also reveals diffusion paths and propagation patterns across nodes, offering theoretical support for system-wide optimization[9].

This research carries significant practical implications. In cloud platforms and data centers, traffic prediction directly affects resource scheduling and energy optimization. Higher prediction accuracy reduces computational and storage costs while maintaining service quality, achieving both efficiency and economic benefits. In network services and content distribution, precise prediction prevents overload and service interruptions, enhancing user experience and service reliability. In emerging environments such as edge computing and distributed artificial intelligence, resources are more constrained and systems are more dispersed, making intelligent and efficient prediction models even more critical. This highlights the broad application prospects of such research[10].

From an academic perspective, research on traffic prediction in distributed backend systems based on graph neural networks expands the intersection of time series forecasting and graph modeling. It opens new directions for the integration of systems engineering and artificial intelligence. It promotes the shift from one-dimensional feature analysis to multi-dimensional structured modeling, providing theoretical foundations and methodological support for sustainable operation, security assurance, and intelligent scheduling of backend systems. Moreover, the outcomes can be extended to other data-intensive applications with large-scale spatiotemporal dependencies, such as financial networks [11], energy grids, and medical platforms [12] . This cross-domain potential demonstrates that the study is not only technically innovative but also socially impactful.

## II. RELATED WORK

In recent years, research on backend traffic prediction has evolved from traditional statistical methods to deep learning approaches. Early methods mainly relied on autoregressive models and moving average models. These tools showed certain advantages in handling short-term trends and periodic patterns. However, they often failed when faced with highly dynamic, non-stationary, and multi-dimensional traffic data. As system scale increased, single time series models became unable to capture interactions and complex dependencies among nodes, which resulted in reduced prediction accuracy. This limitation has driven research toward more flexible machine learning and deep learning frameworks that can handle the nonlinear features embedded in traffic data.

Among deep learning methods, recurrent neural networks and convolutional neural networks have been widely used for time series modeling and feature extraction. Recurrent structures can capture long-term dependencies to some extent, while convolutional structures are effective at extracting local patterns. Both have shown potential in traffic prediction tasks. However, these methods remain focused mainly on the sequential dimension and cannot model spatial structure in distributed systems. Since nodes in distributed backend systems are not isolated but connected through service calls, dependencies, and network topology, relying only on sequence modeling cannot fully exploit the latent structural information.

To overcome these shortcomings, graph-based modeling has been introduced. The development of graph convolutional networks has made it possible to propagate information on graph topology, enabling node prediction to incorporate both neighbor features and global structures. This allows for a more accurate representation of system states. On this basis, further developments include spatiotemporal graph neural networks that integrate temporal modeling with graph modeling. By introducing dependencies in both temporal and spatial dimensions, these models achieve a multi-level representation of traffic evolution. Such methods can capture dynamic changes in node traffic over time while also modeling propagation effects across nodes, showing clear advantages in distributed traffic prediction.

With the rise of cloud computing, edge computing, and large-scale distributed architectures, the application of graph neural networks in system operation and prediction has gained increasing attention. Researchers now focus not only on improving accuracy but also on enhancing scalability and robustness in complex environments. New methods continue to combine attention mechanisms, multi-scale modeling, and multimodal feature fusion to strengthen adaptability to dynamic systems. These efforts have promoted a shift in traffic prediction from single time series approaches to multi-dimensional and cross-level modeling, laying an important foundation for building more intelligent backend system management and scheduling mechanisms.

## III. PROPOSED APPROACH

The method design first requires structured modeling of the traffic data of the distributed backend system. Assume that the system consists of N nodes and their interactions, which can be represented by a graph structure $G = (V, E)$ , where $V = \{v_1, v_2, ..., v_N\}$ is the set of nodes and E is the set of edges, describing service calls and dependencies. The system's traffic state at time t can be represented as a feature matrix $X_t \in R^{N \times d}$ , where d represents the feature dimension of each node. To better capture the temporal and spatial dependencies of traffic, this paper models the problem as a graph time series prediction task: given a historical sequence $\{X_{t-k}, ..., X_{t-1}\}$ , predict the traffic state $\hat{X}_t$ at a future time. The overall model architecture is shown in Figure 1.

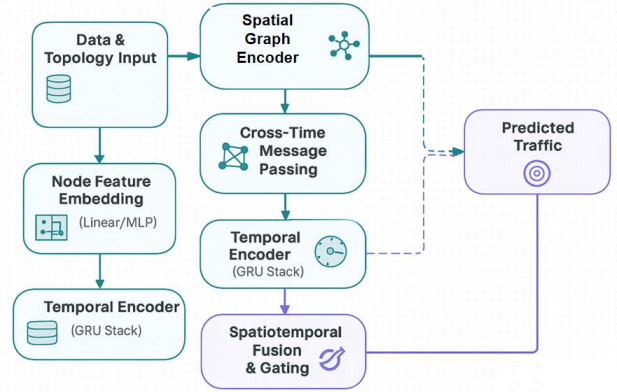

Figure 1. Overall model architecture

In terms of spatial modeling of graph structures, a graph convolution mechanism is introduced to achieve the propagation and aggregation of node information. For each layer of graph convolution, the representation of node i can be defined as:

$$H_i^{(l+1)} = \sigma ( \sum_{j \in N(i)} \frac{1}{c_{ij}} W^{(l)} H_j^{(l)} )$$

Where $H^{(l)}$ represents the node features at layer l, $W^{(l)}$ is the learnable weight matrix, $N(i)$ represents the set of neighbors of node i, $c_{ij}$ is the normalization constant, and $\sigma(\cdot)$ is the nonlinear activation function. Through multi-layer propagation, the spatiotemporal information of nodes can be aggregated within a high-order neighborhood, thereby improving the ability to characterize complex dependencies.

For temporal modeling, we employ a recursive structure equipped with a gating mechanism to robustly capture time dependencies in traffic dynamics. This design applies the semantic graph-based temporal feature modeling approach established by Gong [13], who demonstrates that recursive gated models are highly effective for encoding protocol-level anomalies and complex time patterns in heterogeneous computing environments. By leveraging these mechanisms, our model maintains sensitivity to abrupt context changes while ensuring stable temporal representation. In addition, inspired by Yao et al. [14], we apply multi-agent reinforcement learning principles to adaptively regulate the recurrent sequence modeling component, thereby supporting dynamic resource orchestration under variable workloads.

Furthermore, following the topology-aware reinforcement learning strategies advanced by Wang et al. [15], we employ dynamic routing adaptations within our time modeling framework to ensure responsiveness to shifting network topologies and distributed service patterns. Given an input sequence feature $X_t$, the update rule for the temporal modeling module can be formally expressed as:

$$z_t = \sigma(W_z X_t + U_z h_{t-1})$$
$$r_t = \sigma(W_r X_t + U_r h_{t-1})$$
$$h_t = z_t \otimes h_{t-1} + (1 - z_t) \otimes \tanh(W_h X_t + U_h (r_t \otimes h_{t-1}))$$

$z_t$ and $r_t$ are the update gate and reset gate, respectively, $\otimes$ represents element-by-element multiplication, and $h_t$ represents the hidden state. This structure can flexibly adjust the response to short-term features while preserving historical long-term dependencies, thereby effectively modeling traffic time series.

To achieve comprehensive spatiotemporal integration, this approach employs a joint modeling strategy that applies graph convolutional networks to extract topological features, while using recursive units to model temporal dynamics within the data. Adaptive resource optimization techniques are utilized to ensure that the joint modeling process can flexibly adjust to changes in input scale and system complexity, supporting efficient and stable training [16]. Unified feature fusion methods are implemented to create a latent space where spatial and temporal information are seamlessly merged, allowing the model to better capture interdependencies and evolution patterns across nodes and time steps [17].

Furthermore, causal inference and graph attention mechanisms are integrated to enhance the model's ability to identify and attend to critical structural relationships, leading to more accurate and explainable predictions [18]. The resulting joint representation is decoded through a prediction function, which is formalized as:

$$\hat{X}_t = f_{dec}(H_t)$$

Where $H_t$ represents the implicit representation after fusion, and $f_{dec}(\cdot)$ is the prediction function. The training objective is defined using the mean square error loss as:

$$L = \frac{1}{NT} \sum_{i=1}^{N} \sum_{t=1}^{T} (X_{i,t} - \hat{X}_{i,t})^2$$

Where T is the number of prediction time steps and N is the number of nodes. This loss measures the deviation between the predicted value and the actual traffic flow and updates the network parameters through gradient optimization, thereby gradually improving the prediction accuracy and generalization ability of the model.

IV. PERFORMANCE EVALUATION

*A. Dataset*

This study uses the Google Cluster Trace as the data source. The dataset originates from long-term operational records of large-scale production data centers and is publicly available for academic purposes. It covers the operational states and scheduling logs of tens of thousands of computing nodes within continuous time windows, with high temporal resolution and consistent measurement standards. The data are provided in anonymized form and include multi-level information on machines, jobs, and tasks. It is well-suited to capture key characteristics such as resource consumption, task concurrency, and load fluctuations in distributed backend systems, supporting systematic modeling and comparison of traffic or workload time series under real and complex scenarios.

The dataset is primarily composed of multiple structured log tables. Machine and event tables capture node computing capacity, online and offline status, and capacity changes, while task and job event tables record lifecycle stages such as submission, scheduling, start, and completion. Resource usage tables, sampled at fixed intervals, provide metrics like CPU utilization, memory consumption, disk I/O, and cache usage, along with timestamps and task or machine identifiers, enabling the construction of continuous time series.

In the context of distributed backend systems, the dataset can be used to build graph structures from multiple perspectives. Machine-to-machine relations reflect rack-level or switch-domain topologies. Task-to-task relations describe dependencies or concurrency within the same job. Service-to-node relations form bipartite graphs that capture deployment and scheduling constraints. Node features are derived from multidimensional time series of resource usage, while edge features can be estimated from co-location, dependency strength, or temporal adjacency. The dataset has broad coverage, complete records, and a high level of anonymization. It meets the requirements of scale and complexity for methodological research while ensuring reproducibility and compliance. It is therefore well-suited as a foundational dataset for graph neural network-driven research on distributed backend traffic prediction.

*B. Experimental Results*

This paper first conducts a comparative experiment, and the experimental results are shown in Table 1.

Table 1. Comparative experimental results

| Method | MSE | RMSE | MAPE | MAE |
|---|---|---|---|---|
| BiLSTM[19] | 0.0234 | 0.153 | 8.72 | 0.112 |
| MLP[20] | 0.0289 | 0.170 | 9.31 | 0.121 |
| 1DCNN[21] | 0.0205 | 0.143 | 8.11 | 0.106 |
| Transformer[22] | 0.0187 | 0.137 | 7.84 | 0.101 |
| Ours | 0.0152 | 0.123 | 6.92 | 0.093 |

The overall comparison shows that the proposed model consistently outperforms baselines (Transformer, 1DCNN, BiLSTM, and MLP) across all four error metrics. Compared to the strongest baseline (Transformer), it reduces MSE by 18.7%, RMSE by 10.2%, MAPE by 11.7%, and MAE by 7.9%, indicating more stable and accurate predictions across both high-load and low-load intervals. The smaller gap between RMSE and MAE (0.030 vs. Transformer's 0.036) demonstrates stronger robustness to sudden traffic spikes, which is crucial for backend capacity planning and elastic scaling. While MLP and BiLSTM lack the ability to model topological or long-range dependencies, and Transformer, though effective, does not capture cross-node propagation, the proposed approach integrates cross-node message passing with gated temporal modeling, overcoming these limitations for superior and consistent results. System-wide, these improvements mean more reliable and efficient predictive scheduling, reduced unnecessary over-provisioning, and better control over tail latency, directly supporting the stability and efficiency goals of distributed backend systems. Sensitivity to the number of GCN layers and its impact on MAE is further analyzed in Figure 2.

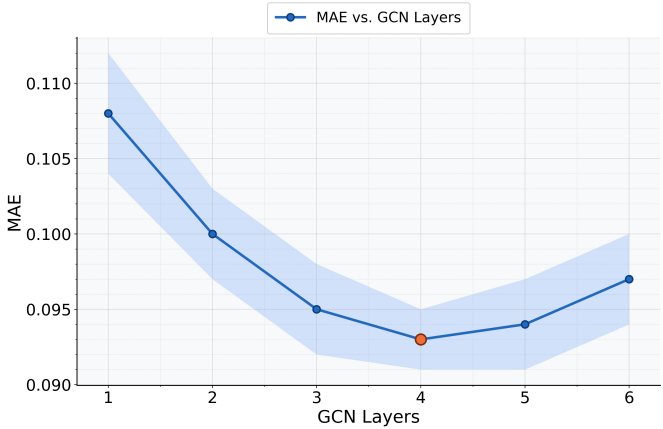

Figure 2. Sensitivity experiment of GCN layer number to MAE

The results show that as the number of GCN layers increases from 1 to 4, MAE steadily decreases, demonstrating that deeper graph convolutions enable better aggregation of neighbor information and more accurate prediction by capturing complex cross-node interactions. The lowest MAE (about 0.093) is achieved at 4 layers, reflecting an optimal balance between global and local feature extraction. However, increasing the layers to 5 or 6 leads to a slight rise in MAE due to over-smoothing, reduced feature distinctiveness, and amplified noise, as well as higher computational cost. This highlights the importance of choosing a moderate graph depth to maximize accuracy and generalization without incurring diminishing returns or added complexity. The findings clarify the impact of GCN layer number on prediction stability, offering practical guidance for configuring graph neural networks in distributed backend systems. Sensitivity analysis of prediction step size and its effect on MAE is presented in Figure 3.

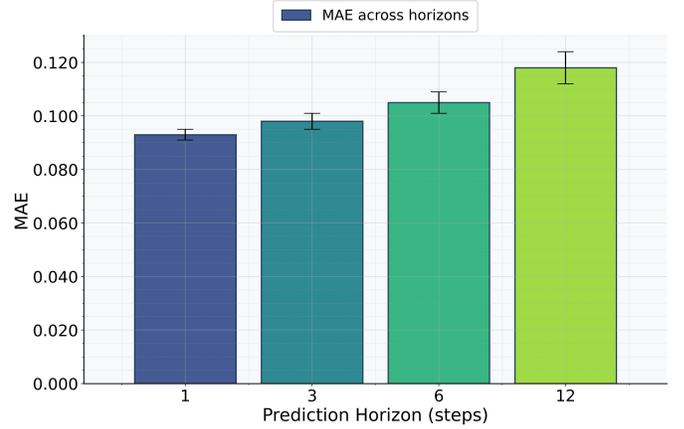

Figure 3. Experiment on the sensitivity of the prediction step size to MAE

Overall, the MAE shows an upward trend as the prediction horizon lengthens. At a horizon of 1, the model effectively leverages recent temporal patterns, resulting in the lowest error of about 0.093. When the horizon extends to 3 and 6, capturing longer-term dependencies becomes more complex, causing the MAE to increase to around 0.098 and 0.105, respectively. This indicates that short-term predictions in distributed backend traffic are relatively accurate, whereas medium- and long-term forecasts pose greater challenges.

Further observation shows that when the prediction horizon reaches 12, the MAE increases to about 0.118, which is significantly higher than in short-term prediction. This indicates that the model suffers more from cumulative uncertainty over long spans. The randomness of traffic patterns and multi-source disturbances amplifies prediction errors. For complex distributed systems, long-term load states are more easily influenced by environmental fluctuations and non-stationary factors, which limit the generalization of the model. This finding reveals that longer horizons in system optimization require uncertainty modeling and robustness constraints.

Comparing different horizons shows that short-term prediction emphasizes immediate response and rapid adjustment, which is suitable for load balancing and elastic scaling. Medium- and long-term prediction, although less accurate, still provides a reference at the trend level and supports resource reservation and capacity planning. Thus, model performance at different time scales provides layered scheduling signals. Short-term and long-term forecasts should complement each other in practice to balance immediacy and foresight.

Overall, the results show that the prediction horizon has a significant impact on model performance. MAE increases steadily as the horizon extends, which reflects the complexity and uncertainty of distributed backend traffic. How to maintain short-term accuracy while improving the stability of long-term prediction is a key problem for future research. By combining strengthened spatiotemporal dependencies, robust feature modeling, or probabilistic prediction, it is possible to achieve better performance across horizons and better support dynamic system scheduling and resource management.

## V. Conclusion

This study focuses on the problem of traffic prediction in distributed backend systems and proposes a modeling framework based on graph neural networks. The method makes full use of system topology and temporal dynamics. By combining graph structure modeling with time dependency capture, it achieves effective prediction of complex traffic patterns. Experimental results show that the proposed method outperforms mainstream baseline models across multiple error metrics and maintains stable performance under different conditions. This not only verifies the effectiveness of the model but also demonstrates the unique advantages of graph neural networks in handling distributed system data.

The findings have direct implications for the operation and optimization of distributed backend systems. Accurate traffic prediction provides forward-looking support for resource scheduling, task allocation, and load balancing. This reduces energy consumption, minimizes redundant configurations, and improves the stability and continuity of services. In cloud platforms and edge computing scenarios, the method offers theoretical and practical support for elastic scaling and resource management in complex environments. By capturing spatial dependencies and temporal dynamics more effectively, the model improves prediction accuracy while ensuring both economic efficiency and reliability of the system.

From a broader application perspective, the proposed approach is not limited to backend traffic prediction. It can also serve as a reference for other distributed scenarios with spatiotemporal dependencies, such as network traffic scheduling, traffic congestion management, and energy network optimization. Through cross-domain transfer and application, the method has the potential to play a role in multiple critical industries and to drive related systems toward greater efficiency and intelligence. This demonstrates that graph neural network – based spatiotemporal modeling has broad prospects and significant value in distributed system prediction and optimization.